\documentclass[prologue]{lipics-v2021}

\usepackage{amsmath, amssymb, mathrsfs, tikz, tikz-cd, mathpazo, anyfontsize, xargs, environ, tabularx, booktabs, makecell, colortbl, float, art.cls/colorpal, art.cls/ct, art.cls/sset, art.cls/lim, art.cls/joinargs}
\usepackage[utf8]{inputenc}
\usepackage[OT1]{fontenc}
\usepackage[prefix=bonak]{art.cls/xkeymask}

\usetikzlibrary{patterns}

\makeatletter
\define@cmdkey[bonak]{X}{D}{(#1)}

\define@cmdkey[bonak]{frame}{D}{(#1)}
\define@cmdkey[bonak]{layer}{D}{(#1)}
\define@cmdkey[bonak]{layer}{d}{(#1)}
\define@cmdkey[bonak]{painting}{D}{(#1)}
\define@cmdkey[bonak]{painting}{E}{(#1)}
\define@cmdkey[bonak]{painting}{d}{(#1)}
\define@cmdkey[bonak]{fullframe}{D}{(#1)}

\define@cmdkey[bonak]{restrframe}{D}{(#1)}
\define@cmdkey[bonak]{restrframe}{d}{(#1)}
\define@cmdkey[bonak]{restrlayer}{D}{(#1)}
\define@cmdkey[bonak]{restrlayer}{d}{(#1)}
\define@cmdkey[bonak]{restrlayer}{l}{(#1)}
\define@cmdkey[bonak]{restrpainting}{D}{(#1)}
\define@cmdkey[bonak]{restrpainting}{E}{(#1)}
\define@cmdkey[bonak]{restrpainting}{d}{(#1)}
\define@cmdkey[bonak]{restrpainting}{c}{(#1)}

\define@cmdkey[bonak]{cohframe}{D}{(#1)}
\define@cmdkey[bonak]{cohframe}{d}{(#1)}
\define@cmdkey[bonak]{cohlayer}{D}{(#1)}
\define@cmdkey[bonak]{cohlayer}{d}{(#1)}
\define@cmdkey[bonak]{cohlayer}{l}{(#1)}
\define@cmdkey[bonak]{cohpainting}{D}{(#1)}
\define@cmdkey[bonak]{cohpainting}{E}{(#1)}
\define@cmdkey[bonak]{cohpainting}{d}{(#1)}
\define@cmdkey[bonak]{cohpainting}{c}{(#1)}
\makeatother

\newcommandx{\X}[3][1,2,3]{
  \ensuremath{{\color{carolina}{\nu\textsf{Set}}}_{#1}^{#2}}
  \setkeys*[bonak]{X}{#3}
}

\newcommandx{\prim}[6][2,3,4,5,6]{
  \ensuremath{\mathsf{\color{indian-yellow}{#1}}_{#2}^{
    \joinargs[#3][#4][#5]}}
  \setkeys*[bonak]{#1}{#6}
}

\newcommandx{\restr}[8][2,3,4,5,6,7,8]{
  \ensuremath{\mathsf{\color{russian-green}{restr}}_{
    \joinargs[\mathsf{\color{indian-yellow}{#1}}][#2][#3][#4]}^{\joinargs[#5][#6][#7]}}
  \setkeys*[bonak]{restr#1}{#8}
}

\newcommandx{\depsmacro}[5][4,5]{
  \ensuremath{\mathsf{\color{strawberry}{#1}}_{
    \joinargs[\mathsf{\color{#3}{#2}}]}^{\joinargs[#4][#5]}}
}

\newcommandx{\coh}[9][2,3,4,5,6,7,8,9]{
  \ensuremath{\mathsf{\color{chestnut}{coh}}_{
    \joinargs[\mathsf{\color{indian-yellow}{#1}}][#2][#3][#4][#5][#6]}^{\joinargs[#7][#8]}}
  \setkeys*[bonak]{coh#1}{#9}
}

\newcommandx{\cohtwo}[9][2,3,4,5,6,7,8,9]{
  \ensuremath{\mathsf{\color{chestnut}{coh2}}_{
    \joinargs[\mathsf{\color{indian-yellow}{#1}}][#2][#3][#4][#5][#6]}^{\joinargs[#7][#8]}}
  \setkeys*[bonak]{coh#1}{#9}
}

\newcommandx{\HSet}[1][1=]{\ensuremath{\mathsf{\color{spanish-blue}{HSet}}_{#1}}}
\newcommand{\HProp}{\ensuremath{\mathsf{\color{spanish-blue}{HProp}}}}
\newcommandx{\HGpd}[1][1=]{\ensuremath{\mathsf{\color{spanish-blue}{HGpd}}_{#1}}}

\newcommand{\unittype}{\ensuremath{\mathsf{unit}}}

\newcommand{\unitpoint}{\ensuremath{\ast}}

\newcommand{\defeq}{\ensuremath{\triangleq}}
\newcommand{\refl}{\ensuremath{\mathsf{refl}}}

\newcommand{\imp}{\rightarrow}

\newcommand{\ap}{\mathsf{ap}\;}
\renewcommand{\D}{D}
\newcommand{\hdD}{D.1}
\newcommand{\tlD}{D.2}
\renewcommand{\d}{d}
\renewcommand{\E}{E}
\renewcommand{\l}{l}
\renewcommand{\c}{c}
\newcommand{\pair}[2]{#1, #2}

\newcommand{\UIP}{\textsf{UIP}}

\newcommand{\udensdash}[1]{%
    \tikz[baseline=(todotted.base)]{
        \node[inner sep=1pt,outer sep=0pt] (todotted) {$#1$};
        \draw[densely dashed] (todotted.south west) -- (todotted.south east);
    }%
}%

\newcommand{\smfontsize}{\fontsize{8}{11}\selectfont}

\newcommandx{\framep}[2][1,2]{\prim{frame}[][#1][#2][][]}
\newcommandx{\layer}[2][1,2]{\prim{layer}[][#1][#2][][]}
\newcommandx{\painting}[2][1,2]{\prim{painting}[][#1][#2][][]}
\newcommandx{\restrf}[4][1,2,3,4]{\restr{frame}[][#3][#4][#1][#2][][]}
\newcommandx{\restrl}[4][1,2,3,4]{\restr{layer}[][#3][#4][#1][#2][][]}
\newcommandx{\restrp}[4][1,2,3,4]{\restr{painting}[][#3][#4][#1][#2][][]}
\newcommandx{\cohf}{\coh{frame}[][][][][][][][]}
\newcommandx{\cohl}{\coh{layer}[][][][][][][][]}
\newcommandx{\cohp}{\coh{painting}[][][][][][][][]}
\newcommandx{\fullframe}[1][1]{\prim{fullframe}[][#1][][][]}

\newcommandx{\frametype}[2][1,2]{\prim{FRAME}[][#1][#2][][]}
\newcommandx{\layertype}[2][1,2]{\prim{LAYER}[][#1][#2][][]}
\newcommandx{\paintingtype}[2][1,2]{\prim{PAINTING}[][#1][#2][][]}
\newcommandx{\restrftype}[2][1,2]{\restr{FRAME}[][][][#1][#2][][]}
\newcommandx{\restrltype}[2][1,2]{\restr{LAYER}[][][][#1][#2][][]}
\newcommandx{\restrptype}[2][1,2]{\restr{PAINTING}[][][][#1][#2][][]}
\newcommandx{\cohftype}[2][1,2]{\coh{FRAME}[][][][][][#1][#2][]}
\newcommandx{\cohltype}[2][1,2]{\coh{LAYER}[][][][][][#1][#2][]}
\newcommandx{\cohptype}[2][1,2]{\coh{PAINTING}[][][][][][#1][#2][]}
\newcommandx{\cohttype}[2][1,2]{\cohtwo{FRAME}[][][][][][#1][#2][]}

\newcommandx{\deps}[2][1,2]{\depsmacro{deps}{restr}{russian-green}[#1][#2]}
\newcommandx{\fulldeps}[1][1]{\depsmacro{deps}{fullrestr}{russian-green}[#1]}
\newcommandx{\depscohs}[2][1,2]{\depsmacro{deps}{coh}{chestnut}[#1][#2]}
\newcommandx{\fulldepscohs}[1][1]{\depsmacro{deps}{fullcoh}{chestnut}[#1]}
\newcommandx{\depscoht}[2][1,2]{\depsmacro{deps}{coh2}{chestnut}[#1][#2]}
\newcommandx{\fulldepscoht}[1][1]{\depsmacro{deps}{fullcoh2}{chestnut}[#1]}
\newcommandx{\depstype}[2][1,2]{\depsmacro{DEPS}{restr}{russian-green}[#1][#2]}
\newcommandx{\fulldepstype}[1][1]{\depsmacro{DEPS}{fullrestr}{russian-green}[#1]}
\newcommandx{\depscohstype}[2][1,2]{\depsmacro{DEPS}{coh}{chestnut}[#1][#2]}
\newcommandx{\fulldepscohstype}[1][1]{\depsmacro{DEPS}{fullcoh}{chestnut}[#1]}
\newcommandx{\depscohttype}[2][1,2]{\depsmacro{DEPS}{coh2}{chestnut}[#1][#2]}
\newcommandx{\fulldepscohttype}[1][1]{\depsmacro{DEPS}{fullcoh2}{chestnut}[#1]}

\newcommandx{\coht}[7][1,2,3,4,5,6,7]{\cohtwo{frame}[#3][#4][#5][#6][#7][#1][#2][]}

\newcommand{\kstar}{{\star}}

\newcommand{\linkicon}{\texorpdfstring{\faLink}{🔗}}

\newcolumntype{Y}{>{\centering\arraybackslash}X}
\NewEnviron{eqntable}[1]{
  \begin{table}[H]
  \smfontsize
  \begin{tabularx}{\linewidth}{
    @{}
    >{$}l<{$}
    >{$}c<{$}
    >{$}c<{$}
    >{$}Y<{$}
    @{}}
    \toprule
    \BODY
    \bottomrule
  \end{tabularx}
  \caption{#1}
  \end{table}
}

\NewDocumentCommand{\eqnline}{m m m m}{#1 & #2 & #3 & #4 \\}
\newcommandx*{\mc}[1]{\multicolumn{4}{c}{\emph{#1}} \\\\}

\newcommandx*{\eqnarg}[3]{\ifinmask[bonak]{#1}[#2]{\{#2:#3\}}{(#2:#3)}}

\title{The very dependent recursive structure of iterated parametricity in indexed form}
\author{Hugo Herbelin}{Université Paris Cité, Inria, CNRS, IRIF, Paris}{Hugo.Herbelin@inria.fr}{}{}
\author{Ramkumar Ramachandra}{Unaffiliated}{r@artagnon.com}{}{}
\authorrunning{H. Herbelin and R. Ramachandra}
\Copyright{Hugo Herbelin and Ramkumar Ramachandra}
\nolinenumbers

\begin{document}
\maketitle
\begin{abstract}
  Reynolds' parametricity originally equips types with proof-irrelevant binary propositional relations over the types. But such relations can also be taken proof-relevant or unary, and described either in an indexed or fibred way. Parametricity can be iterated, and when types are sets, this results in an interpretation of sets as augmented simplicial sets in the unary case, or cubical sets in the binary case.

  In earlier work, equations were given describing the $n$-ary iterated parametricity translation of sets in indexed form. The construction was formalised in Rocq by induction on a large structure embedding equational reasoning.

  The current work analyses the dependency structure of the earlier work leading to a presentation of the construction replacing equational reasoning with definitional reasoning. The new construction is very dependent, based on an induction that requires interleaving the specification of the induction hypothesis and the construction of the induction step. At the same time, the construction reduces to its computational essence and can be described in full detail, closely following the new machine-checked formalisation.
\end{abstract}

\section{Introduction}

\paragraph*{Parametricity and its ramifications}

Introduced by Reynolds in the context of programming language semantics~\cite{reynolds83}, relational parametricity is connected to several similar concepts in logic and mathematics. In its original formulation, it associates binary propositional relations to types. The variant in~\cite{GhaniNordvallOrsanigo16} associates proof-relevant binary relations to types, and the variant in~\cite{bernardy10} associates relations of arbitrary arity to types. This includes the case of unary relations which assigns predicates to types in the proof-irrelevant case, and families to types in the proof-relevant case. The unary relation case expresses what it means to belong in a type, and the binary relation case expresses what it means to be observationally equal in a type. The canonical predicate associated to $A \rightarrow B$ is the predicate over functions $f$ in $A \rightarrow B$ such that, for all $a$ in the canonical predicate associated to $A$, $f(a)$ is in the canonical predicate associated to $B$. Similarly, the canonical relation associated to $A \rightarrow B$ is the relation over functions $f$ and $f'$ in $A \rightarrow B$ such that, for all $a$ and $a'$ related through the canonical relation associated to $A$, $f(a)$ and $f'(a')$ are related through the canonical relation associated to $B$.

Unary parametricity over a language is an extreme form of realisability using the language itself as language of realisers. As such, it relates to \emph{reducibility}, also known as \emph{Tait computability}~\cite{Tait67}, which is a semantic form of realisability. In the same vein, binary parametricity over a language can be seen as defining \emph{logical relations}~\cite{Plotkin73} over itself, since logical relations are a binary form of reducibility.

Parametricity, like realisability, reducibility, and logical relations, is commonly expressed in \emph{indexed} form. Let us use a type-theoretic language featuring a type of subsingleton types called $\HProp$ and a type of sets $\HSet$. Then, predicates have type $A \rightarrow \HProp$, families have type $A \rightarrow \HSet$, relations have type $A \times A \rightarrow \HProp$, and relevant relations, sometimes called \emph{correspondences}, have type $A \times A \rightarrow \HSet$. All constructions in indexed form can also be expressed in the \emph{fibred} form, via the indexed-fibred correspondence: there is an equivalence between $A \rightarrow \HProp$ and $\Sigma B: \HSet. (B \stackrel{inj}{\rightarrow} A)$ in the proof-irrelevant case, and between $A \rightarrow \HSet$ and $\Sigma B: \HSet. (B \rightarrow A)$ in the proof-relevant case, for $A: \HSet$. A fibred approach to parametricity was considered in~\cite{atkey14}, who also suggested to iterate parametricity.

\paragraph*{Iterated parametricity}

The internalisation of parametricity in unary indexed form was studied in~\cite{bernardy12} leading to a type theory of iterated parametricity. Such a type theory can be interpreted in the ``unary'' form of cubical sets~\cite{bernardy15}. Cubical sets, in their standard definition as presheaves over a cubical category, can be seen as living on the fibred side of iterated parametricity.\footnote{Interestingly, this contrasts with the case of Grothendieck construction where presheaves live on the indexed side and fibrations on the fibred side} This was particularly emphasised in~\cite{johann17} which interprets cubical sets as higher-dimensional parametricity in fibred form. While iterated binary parametricity produces semi-cubical sets, it was also observed that iterated unary parametricity produces augmented semi-simplicial sets.\footnote{The authors learned it from Hugo Moeneclaey and Thorsten Altenkirch}

\begin{table}[H]
  \begin{tabularx}{\linewidth}{cYcYYc}
    \toprule
    \makecell{arity}  & \makecell{style}   & \makecell{iteration}    & \makecell{relevance}        & \makecell{logical structure}   \\
    \midrule
    \makecell{unary}  & \makecell{indexed} & \makecell{non-iterated} & \makecell{proof-irrelevant} & \makecell{predicate}           \\
                      &                    & \makecell{non-iterated} & \makecell{relevant}         & \makecell{family}              \\
                      &                    & \makecell{iterated}     & \makecell{relevant}         & \makecell{\emph{this work}}    \\
    \midrule
                      & \makecell{fibred}  & \makecell{non-iterated} & \makecell{proof-irrelevant} & \makecell{subobject}           \\
                      &                    & \makecell{non-iterated} & \makecell{relevant}         & \makecell{fibration}           \\
                      &                    & \makecell{iterated}     & \makecell{relevant}         & \makecell{augmented            \\ semi-simplicial set}   \\
    \midrule
    \makecell{binary} & \makecell{indexed} & \makecell{non-iterated} & \makecell{proof-irrelevant} & \makecell{relation}            \\
                      &                    & \makecell{non-iterated} & \makecell{relevant}         & \makecell{correspondence}      \\
                      &                    & \makecell{iterated}     & \makecell{relevant}         & \makecell{\emph{this work}}    \\
    \midrule
                      & \makecell{fibred}  & \makecell{non-iterated} & \makecell{proof-irrelevant} & \makecell{fibred relation}     \\
                      &                    & \makecell{non-iterated} & \makecell{relevant}         & span                         & \\
                      &                    & \makecell{iterated}     & \makecell{relevant}         & \makecell{semi-cubical set}    \\
    \bottomrule
  \end{tabularx}
\end{table}

\paragraph*{Iterated parametricity in indexed form}

An informal description of semi-simplicial sets in indexed form was given in~\cite{UFwiki2012}. Several formal definitions in type theory were later proposed~\cite{voevodsky12,herbelin15,part15,altenkirch16}. In the previous work, a notion of $\nu$-sets generalising augmented semi-simplicial and semi-cubical sets and reflecting $\nu$-ary iterated parametricity was given~\cite{hr25}, together with a construction of $\nu$-sets in \emph{indexed form}. As an example, for $\nu=2$, instead of semi-cubical sets in fibred form, a family of relevant relations was used.

\begin{equation}
  \tag{fibred}
  \begin{tikzcd}
    X_0: \HSet & X_1: \HSet \arrow[l, "\partial^L" description, shift left=2] \arrow[l, "\partial^R" description, shift right=2] & X_2: \HSet \arrow[l, "\partial^{L\kstar}" description, shift left=6] \arrow[l, "\partial^{R\kstar}" description, shift left=2] \arrow[l, "\partial^{\kstar L}" description, shift right=2] \arrow[l, "\partial^{\kstar R}" description, shift right=6] & \ldots
  \end{tikzcd}
\end{equation}

\begin{equation}
  \tag{relevant relations}
  \begin{array}{lll}
    E_0 & :               \HSet                                                                          \\
    E_1 & :               E_0 \times E_0 \rightarrow  \HSet                                              \\
    E_2 & : \Pi a b c d.\,  E_1(a,b) \times E_1 (c,d) \times E_1(a,c) \times E_1 (b,d) \rightarrow \HSet \\
    \ldots
  \end{array}
\end{equation}

%
\appendmask[bonak]{layer}[D]
\appendmask[bonak]{painting}[D]
\appendmask[bonak]{restrframe}[D]
\appendmask[bonak]{restrlayer}[D, d]
\appendmask[bonak]{restrpainting}[D, d]
\appendmask[bonak]{cohframe}[D]
\appendmask[bonak]{cohlayer}[D, d]
\appendmask[bonak]{cohpainting}[D, d]

\newcommandx{\Xp}[1]{\X[#1][][]}
\newcommandx{\Xto}[3][3=]{\X[#1][<#2][#3]}
\newcommandx{\Xcomp}[3][3=]{\X[#1][=#2][#3]}
\newcommandx{\Xfrom}[3][3=]{\X[#1][\geq#2][#3]}

\renewcommandx{\framep}[5][1,2,3,4,5]{\prim{frame}[][#2][#3][#4][#5]}
\renewcommandx{\layer}[5][1,2,3,4,5]{\prim{layer}[][#2][#3][#4][#5]}
\renewcommandx{\painting}[5][1,2,3,4,5]{\prim{painting}[][#2][#3][#4][#5]}

\renewcommandx{\restrf}[7][1,2,3,4,5,6,7]{\restr{frame}[][#2][#3][#4][#5][#6][#7]}
\renewcommandx{\restrl}[7][1,2,3,4,5,6,7]{\restr{layer}[][#2][#3][#4][#5][#6][#7]}
\renewcommandx{\restrp}[7][1,2,3,4,5,6,7]{\restr{painting}[][#2][#3][#4][#5][#6][#7]}

\renewcommandx{\cohf}[9][1,2,3,4,5,6,7,8,9]{\coh{frame}[][#2][#3][][][][][#9]}
\renewcommandx{\cohl}[9][1,2,3,4,5,6,7,8,9]{\coh{layer}[][#2][#3][][][][][#9]}
\renewcommandx{\cohp}[9][1,2,3,4,5,6,7,8,9]{\coh{painting}[][#2][#3][][][][][#9]}

Equations specifying $\nu$-sets in indexed form were given, and the well-foundedness of the construction was by induction on a large structure involving equational reasoning. The current work presents a refined construction which expresses the computational essence of the construction.

\section{High-level description of iterated parametricity in indexed form\label{sec:reminder}}

We recall from the previous work how to formally define a type characterising the sequences $E_0, E_1, \ldots$. The type theory we consider is with $\Sigma$ and $\Pi$ types, Martin-L\"of identity type $=$, universes $U_m$, a unit type $\unittype$, natural numbers, a definitionally proof-irrelevant inequality on natural numbers, streams, and function extensionality. The non-dependent variants of $\Pi$ and $\Sigma$ are written $\rightarrow$ and $\times$ respectively. If $e$ is a proof of $x=y$ and $p$ a proof, we write $\overrightarrow{e}{p}$ for substituting $x$ with $y$ in the statement proved by $p$. The canonical inhabitant of $\unittype$ is written $\kstar$, and the subuniverse of types $T : U_m$ satisfying $\UIP : \Pi xy : T. \Pi pq : x = y. p = q$ is written $\HSet[m]$. The definitional proof-irrelevance of inequality is a priori not essential, and could be replaced with propositional proof-irrelevance up to a few extra transports. Our use of function extensionality comes from representing $n$-tuples as functions, and could be avoided by representing $n$-tuples as lists.

Let us sketch how to define iterated parametricity in indexed form for arity $\nu$. The dependency of $E_{n+1}$ into all $E_p$ for $p \leq n$ makes the family $E$ a \emph{very dependent function}~\cite{Hickey96}. The standard way to express it is as a stream whose $(n+1)$-th element $E_{n+1}$ is dependent on the bundle of all previous elements of the stream. We write $\Xto{m}{n}$ for the bundle of all $E_0, \ldots, E_{n-1}$, and we write $\Xfrom{m}{n}{(D)}$ for the residual stream $E_{n}, E_{n+1}, \ldots$, which is dependent on the bundle $E:\Xto{m}{n}$ of previous $E_0, \ldots, E_{n-1}$. The type of $E_n$, also dependent on a bundle $D:\Xto{m}{n}$, is denoted $\Xcomp{m}{n}[D=\D]$, as shown on tables~\ref{tab:coind} and~\ref{tab:core}. When $n = 0$, $\Xto{m}{0}$ is degenerated, and reduced to $\unittype$. Then, $\Xfrom{m}{0}{(\kstar)}$ represents the type of the full stream.

The type $\Xcomp{m}{n}[D={E_0,\ldots,E_{n-1}}]$ of $E_n$ represents a family. It can be canonically presented into a form $\fullframe(E_0,\ldots,E_{n-1}) \rightarrow \HSet$ where $\fullframe$ collects all the dependencies of $E_n$. It is recursively defined in table~\ref{tab:frames}, using $\framep$, $\layer$ and $\painting$. A $\fullframe$ describes a boundary of a standard form (simplex, cube), which we decompose into $n$ $\layer$, one per dimension, each of them made of $\nu$ $\painting$, where a $\painting$ corresponds to a filled frame. Notice that the type $\layer$ relies on an operator of frame restriction $\restrf$ which is defined in table~\ref{tab:faces} for any $\epsilon$ ranging between $1$ and $\nu$, and this restriction operator is in turn defined using $\restrl$ and $\restrp$.

\def\lab{tab:coind}
\providecommand{\lab}{tab:coind}
\begin{eqntable}{Main definition\label{\lab}}
  \eqnline{\Xp{m}}{}{:}{\HSet[m+1]}
  \eqnline{\Xp{m}}{}{\defeq}{\Xfrom{m}{0}[D=\unitpoint]}
  \midrule
  \eqnline{\Xfrom{m}{n}}{\eqnarg{X}{D}{\Xto{m}{n}}}{:}{\HSet[m+1]}
  \eqnline{\Xfrom{m}{n}}{D}{\defeq}{\Sigma E:\Xcomp{m}{n}[D=\D]. \Xfrom{m}{n+1}[D=\pair{D}{E}]}
\end{eqntable}

\renewcommandx{\fullframe}[3][1,2,3]{\prim{fullframe}[#1][#2][][][#3]}

\def\lab{tab:core}
\providecommand{\lab}{tab:core}
\begin{eqntable}{Truncated $\nu$-sets, the core\label{\lab}}
  \eqnline{\Xto{m}{n}}{}{:}{\HSet[m+1]}
  \eqnline{\Xto{m}{0}}{}{\defeq}{\unittype}
  \eqnline{\Xto{m}{n'+1}}{}{\defeq}{\ensuremath{\Sigma}D:\Xto{m}{n'}.\,\Xcomp{m}{n'}[D=\D]}
  \midrule
  \eqnline{\Xcomp{m}{n}}{\eqnarg{X}{D}{\Xto{m}{n}}}{:}{\HSet[m+1]}
  \eqnline{\Xcomp{m}{n}}{D}{\defeq}{\fullframe[m][n][D=\D] \imp \HSet[m]}
\end{eqntable}

\def\lab{tab:frames}
\providecommand{\lab}{tab:frames}
\begin{eqntable}{$\mathsf{frame}$, $\mathsf{layer}$, and $\mathsf{painting}$\label{\lab}}
  \eqnline{\fullframe[m][n]}{\eqnarg{fullframe}{D}{\Xto{m}{n}}}{:}{\HSet[m]}
  \eqnline{\fullframe[m][n]}{D}{\defeq}{\framep[m][n][n][][D=\D]}
  \midrule
  \eqnline{\framep[m][n][p][p \leq n]}{\eqnarg{frame}{D}{\Xto{m}{n}}}{:}{\HSet[m]}
  \eqnline{\framep[m][n][0]}{D}{\defeq}{\unittype}
  \eqnline{\framep[m][n][p+1]}{D}{\defeq}{\Sigma d:\framep[m][n][p][][D=\D].\,\layer[m][n-1][p][][D=\D, d=\d]}
  \midrule
  \eqnline{\layer[m][n][p][p \leq n]}{\makecell{\eqnarg{layer}{D}{\Xto{m}{n+1}} \\ \eqnarg{layer}{d}{\framep[m][n+1][p][][D=\D]}}}{:}{\HSet[m]}
  \eqnline{\layer[m][n][p]}{D~d}{\defeq}{\Pi \epsilon.\painting[m][n][p][][D=\hdD, E=\tlD, d={\restrf[m][p][\epsilon][n][p][][D=\D, d=\d]}]}
  \midrule
  \eqnline{\painting[m][n][p][p \leq n]}{\makecell{\eqnarg{filler}{D}{\Xto{m}{n}} \\ \eqnarg{filler}{E}{\Xcomp{m}{n}[D=\D]} \\ \eqnarg{filler}{d}{\framep[m][n][p][][D=\D]}}}{:}{\HSet[m]}
  \eqnline{\painting[m][n][p][p=n]}{\D~\E~\d}{\defeq}{\E(\d)}
  \eqnline{\painting[m][n][p][p<n]}{\D~\E~\d}{\defeq}{\Sigma l:\layer[m][n-1][p][][D=\D, d=\d].\,\painting[m][n][p+1][][D=\D, E=\E, d=\pair{\d}{\l}]}
\end{eqntable}

\renewcommandx{\cohf}[9][1,2,3,4,5,6,7,8,9]{\coh{frame}[][#2][#3][][][#6,#7][#8][#9]}

\def\lab{tab:faces}
\providecommand{\lab}{fulltab:faces}
\begin{eqntable}{$q$-th projection of $\mathsf{restr}$, or faces\label{\lab}}
  \eqnline{\restrf[m][q][\epsilon][n][p][q \leq n-p]}{\makecell{\eqnarg{restrframe}{D}{\Xto{m}{n+1}} \\ \eqnarg{restrframe}{d}{\framep[m][n+1][p][][D=\D]}}}{:}{\framep[m][n][p][][D=\hdD]}
  \eqnline{\restrf[m][q][\epsilon][n][0]}{D~\unitpoint}{\defeq}{\unitpoint}
  \eqnline{\restrf[m][q][\epsilon][n][p+1]}{D~(\pair{d}{l})}{\defeq}{(\restrf[m][q+1][\epsilon][n][p][][D=\D, d=\d],\restrl[m][q][\epsilon][n-1][p][][D=\D, d=\d, l=\l])}
  \midrule
  \eqnline{\restrl[m][q][\epsilon][n][p][q \leq n-p]}{\makecell{\eqnarg{restrlayer}{D}{\Xto{m}{n+2}} \\ \eqnarg{restrlayer}{d}{\framep[m][n+2][p][][D=\D]} \\ \eqnarg{restrlayer}{l}{\layer[m][n+1][p][][D=\D, d=\d]}}}{:}{\layer[m][n][p][][D=\hdD, d={\restrf[m][q+1][\epsilon][n+1][p][][D=\D, d=\d]}]}
  \eqnline{\restrl[m][q][\epsilon][n][p]}{\D~\d~\l}{\defeq}{\lambda \omega.(\restrp[m][q][\epsilon][n][p][][D=\hdD, E=\tlD, d={\restrf[m][\omega][0][n][p][][D=\D, d=\d]}, c={\l_\omega}])}
  \midrule
  \eqnline{\restrp[m][q][\epsilon][n][p][q \leq n-p]}{\makecell{\eqnarg{restrpainting}{D}{\Xto{m}{n+1}} \\ \eqnarg{restrpainting}{E}{\Xcomp{m}{n+1}[D=\D]} \\\eqnarg{restrpainting}{d}{\framep[m][n+1][p][][D=\D]} \\ \eqnarg{restrpainting}{c}{\painting[m][n+1][p][][D=\D, E=\E, d=\d]}}}{:}{\painting[m][n][p][][D=\hdD, E=\tlD, d={\restrf[m][q][\epsilon][n][p][][D=\D, d=\d]}]}
  \eqnline{\restrp[m][0][\epsilon][n][p][]}{\D~\E~\d~(\pair{l}{\_})}{\defeq}{\l_\epsilon}
  \eqnline{\restrp[m][q+1][\epsilon][n][p][p<n]}{\D~\E~\d~(\pair{l}{c})}{\defeq}{(\restrl[m][q][\epsilon][n-1][p][][D=\D, d=\d, l=\l],\restrp[m][q][\epsilon][n][p+1][][D=\D, E=\E, d=\pair{d}{l}, c=\c])}
\end{eqntable}

\renewcommandx{\restrf}[7][1,2,3,4,5,6,7]{\restr{frame}[][#2][#3][#4][#5][#6][#7]}
\renewcommandx{\restrl}[7][1,2,3,4,5,6,7]{\restr{layer}[][#2][#3][#4][#5][#6][#7]}
\renewcommandx{\restrp}[7][1,2,3,4,5,6,7]{\restr{painting}[][#2][#3][#4][#5][#6][#7]}

\renewcommandx{\cohf}[9][1,2,3,4,5,6,7,8,9]{\coh{frame}[][#2][#3][#4][#5][#6,#7][#8][#9]}
\renewcommandx{\cohl}[9][1,2,3,4,5,6,7,8,9]{\coh{layer}[][#2][#3][#4][#5][#6,#7][#8][#9]}
\renewcommandx{\cohp}[9][1,2,3,4,5,6,7,8,9]{\coh{painting}[][#2][#3][#4][#5][#6,#7][#8][#9]}

\def\lab{tab:coh}
\providecommand{\lab}{fulltab:coh}
\begin{eqntable}{Commutation of $q$-th projection and $r$-th projection, or coherence conditions\label{\lab}}
  \eqnline{\cohf[m][q][r][\epsilon][\omega][n][p][r \leq q \leq n-p]}{\makecell{\eqnarg{cohframe}{D}{\Xto{m}{n+2}} \\ \eqnarg{cohframe}{d}{\framep[m][n+2][p][][D=\D]}}}{:}{\makecell{\restrf[m][q][\epsilon][n][p][][D=\hdD, d={\restrf[m][r][\omega][n+1][p][][D=\D, d=\d]}] \\ = \restrf[m][r][\omega][n][p][][D=\hdD, d={\restrf[m][q+1][\epsilon][n+1][p][][D=\D, d=\d]}]}}
  \eqnline{\cohf[m][q][r][\epsilon][\omega][n][0]}{\D~\unitpoint}{\defeq}{\refl(\unitpoint)}
  \eqnline{\cohf[m][q][r][\epsilon][\omega][n][p+1]}{\D~(\pair{\d}{\l})}{\defeq}{(\cohf[m][q+1][r+1][\epsilon][\omega][n][p][][D=\D, d=\d], \cohl[m][q][r][\epsilon][\omega][n-1][p][][D=\D, d=\d, l=\l])}
  \midrule
  \eqnline{\cohl[m][\epsilon][\omega][q][r][n][p][r \leq q \leq n-p]}{\makecell{\eqnarg{cohlayer}{D}{\Xto{m}{n+3}} \\ \eqnarg{cohlayer}{d}{\framep[m][n+3][p][][D=\D]} \\ \eqnarg{cohlayer}{l}{\layer[m][n+2][p][][D=\D, d=\d]}}}{:}{\makecell{\restrl[m][q][\epsilon][n][p][][D=\hdD, d={\restrf[m][r][\omega][n+1][p][][D=\D, d=\d]}](\restrl[m][r][\omega][n+1][p][][D=\D, d=\d, l=\l]) \\ = \restrl[m][r][\omega][n][p][][D=\hdD, d={\restrf[m][q+1][\epsilon][n+1][p][][D=\D, d=\d]}, l={\restrl[m][q+1][\epsilon][n+1][p][][D=\D, d=\d, l=\l]}]}}
  \eqnline{\cohl[m][\epsilon][\omega][q][r][n][p]}{\D~\d~\l}{\defeq}{\makecell{\lambda \theta .\;\cohp[m][\epsilon][\omega][q][r][n][p][][E=\tlD, c=\l_\theta]}}
  \midrule
  \eqnline{\cohp[m][q][r][\epsilon][\omega][n][p][r \leq q \leq n-p]}{\makecell{\eqnarg{cohpainting}{D}{\Xto{m}{n+2}} \\ \eqnarg{cohpainting}{E}{\Xcomp{m}{n+2}[D=\D]} \\ \eqnarg{cohpainting}{d}{\framep[m][n+2][p][][D=\D]} \\ \eqnarg{cohpainting}{c}{\painting[m][n+2][p][][D=\D, E=\E, d=\d]}}}{:}{\makecell{\restrp[m][q][\epsilon][n][p][][D=\hdD, E=\tlD, d={\restrf[m][r][\omega][n+1][p][][D=\D, d=\d]}, c={\restrp[m][r][\omega][n+1][p][][D=\D, E=\E, d=\d, c=\c]}] \\ = \restrp[m][r][\omega][n][p][][D=\hdD, E=\tlD, D={\restrf[m][q+1][\epsilon][n+1][p][][D=\D, d=\d]}, c={\restrp[m][q+1][\epsilon][n+1][p][][D=\D, E=\E, d=\d, c=\c]}]}}
  \eqnline{\cohp[m][q][0][\epsilon][\omega][n][p][]}{\D~\E~\d~(\pair{\l}{\_})}{\defeq}{\refl(\restrp[m][q][\epsilon][n][p][][D=\hdD, E=\tlD, d={\restrf[m][\omega][0][n][p][][D=\D, d=\d]}, c={\l_\epsilon}])}
  \eqnline{\cohp[m][q+1][r+1][\epsilon][\omega][n][p][p<n]}{\D~\E~\d~(\pair{\l}{\c})}{\defeq}{(\pair{\cohl[m][q][r][\epsilon][\omega][n-1][p][][D=\D, d=\d, l=\l]}{\cohp[m][q][r][\epsilon][\omega][n][p+1][][D=\D, E=\E, d=\pair{d}{l}, c=\c]})}
\end{eqntable}

\renewcommandx{\cohf}{\coh{frame}[][][][][][][][]}
\renewcommandx{\cohl}{\coh{layer}[][][][][][][][]}
\renewcommandx{\cohp}{\coh{painting}[][][][][][][][]}

The definition of $\restrl$ relies on an equality expressing the commutation of the composition of two $\restrf$. The proof of this commutation is made explicit~\footnote{We however omit type conversions in the tables, as if working in extensional type theory. For instance, $\cohf$ is implicitly used in the definition of $\restrl$. For details on where coherence proofs are used, see~\cite{hr25}.} in table~\ref{tab:coh}. It is by recursion on the structure of $\framep$, $\layer$, and $\painting$. This is what $\cohf$ does using the proofs $\cohl$ and $\cohp$. Note that for a fixed $n$, the coherence conditions evaluate to a reflexivity proof, so that the construction builds an effective sequence of types not mentioning coherences anymore.

\renewcommandx{\framep}[2][1,2]{\prim{frame}[][#1][#2][][]}
\renewcommandx{\layer}[2][1,2]{\prim{layer}[][#1][#2][][]}
\renewcommandx{\painting}[2][1,2]{\prim{painting}[][#1][#2][][]}
\renewcommandx{\restrf}[4][1,2,3,4]{\restr{frame}[][#3][#4][#1][#2][][]}
\renewcommandx{\restrl}[4][1,2,3,4]{\restr{layer}[][#3][#4][#1][#2][][]}
\renewcommandx{\restrp}[4][1,2,3,4]{\restr{painting}[][#3][#4][#1][#2][][]}
\renewcommandx{\cohf}[6][1,2,3,4,5,6]{\coh{frame}[][#3][#4][#5][#6][#1][#2][]}
\renewcommandx{\cohl}[6][1,2,3,4,5,6]{\coh{layer}[][#3][#4][#5][#6][#1][#2][]}
\renewcommandx{\cohp}[6][1,2,3,4,5,6]{\coh{painting}[][#3][#4][#5][#6][#1][#2][]}

\section{Analysis of the recursive structure of iterated parametricity in indexed form\label{sec:detailed-construction}}

The well-foundedness of the construction of tables~\ref{tab:coind} to~\ref{tab:coh}, was justified in the previous work by induction on a large structure involving equational reasoning. We show instead in this paper that equational reasoning can be avoided, and that it is enough to build by induction on $n$ the following structure:

\begin{equation*}
  \smfontsize
  \begin{array}{lcl}
    \framep[n][p]   & : & \HSet                                                                                               \\
    \painting[n][p] & : & \framep[n][p] \rightarrow \HSet                                                                     \\
    \restrf[n][p]   & : & \Pi q \leq n-p.\, \Pi \epsilon.\, \framep[n+1][p] \rightarrow \framep[n][p]                         \\
    \restrp[n][p]   & : & \Pi q \leq n-p.\, \Pi \epsilon.\, \Pi d:\framep[n+1][p].\,                                          \\
                    &   & \painting[n+1][p](d) \rightarrow \painting[n][p](\restrf[n][p](d))                                  \\
    \cohf[n][p]     & : & \Pi qr\epsilon\omega.\, \restrf[n][p][q][\epsilon] \circ \restrf[n+1][p][\omega][r] =               \\
                    &   & \restrf[n][p][\omega][r] \circ \restrf[n+1][p][q+1][\epsilon]                                       \\
    \cohp[n][p]     & : & \Pi qr\epsilon\omega.\,\Pi d:\framep[n+2][p].\,                                                     \\
                    &   & \restrp[n][p][q][\epsilon](\restrf[n+1][p][r][\omega](d)) \circ \restrp[n+1][p][r][\omega](d) =     \\
                    &   & \restrp[n][p][r][\omega](\restrf[n+1][p][q+1][\epsilon](d)) \circ \restrp[n+1][p][q+1][\epsilon](d) \\
  \end{array}
\end{equation*}

The price to pay for this conciseness is to mix the specification of the induction hypothesis with the construction of the induction step. On the outset, $\restrf[n][p]$ depends on $\framep[n+1][p]$: it already depends on the ability of building the $\framep$ component of the structure at level $n+1$. There is thus an interdependency between specifying the structure assumed at level $n$ and defining from it the structure at level $n+1$. We have to proceed in a specific order: specifying $\framep[n]$, from which $\painting[n]$ can be specified, from which $\framep[n+1]$ can be built, from which $\restrf[n+1]$ can be specified. Now, the definition of $\framep[n+1][p]$, which proceeds by induction on $p$, itself depends on $\restrf[n][p-1]$ whose type depends in turn on $\framep[n+1][p-1]$. Hence, $\framep[n+1][p]$ and the type of $\restrf[n][p]$ have to be mutually defined.

The definition of $\framep[n+1][p]$ also depends, via $\layer[n][p-1]$, on $\painting[n][p-1]$. Via $\painting[n][p-1]$ and $\restrf[n][p-1]$, it depends also on $\framep[n][p-1]$ by typing dependency. This requires us to assume $\framep[n][p-1]$ and $\painting[n][p-1]$. Moreover, the definition of $\framep[n+1][p]$ depends on $\framep[n+1][p-1]$, so $\framep[n][k]$, $\painting[n][k]$, and $\restrf[n][k]$, for $k < p$, need to be known.

We abbreviate the collection of all $\framep[n][k]$ for $k < p$ as $\framep[n][\lbrack 0,p-1 \rbrack]$ and similarly for $\painting[n][\lbrack 0,p-1 \rbrack]$ and $\restrf[n][\lbrack 0,p-1 \rbrack]$. Conversely, if $\framep[n][\lbrack 0,p \rbrack]$ is a non-empty sequence, we will write $\framep[n][\lbrack 0,p-1 \rbrack]$ for the immediate subsequence of length $p$ and $\framep[n][p]$ for the last component of the sequence. We write $\framep[n]$ for $\framep[n][\lbrack 0,-1 \rbrack]$.

Schematically, using dotted arrows for dependencies in the type and plain arrows for dependencies in the definition, this gives the following graph of dependencies for $p\leq n+1$.

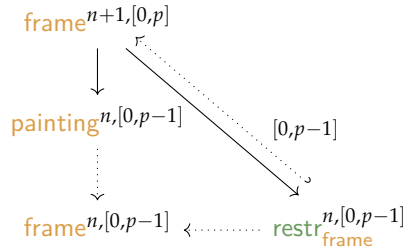
\begin{center}
  \begin{tikzcd}
    \framep[n+1][\lbrack 0,p \rbrack] \arrow[d] \arrow[ddr]\\
    \painting[n][\lbrack 0,p-1 \rbrack] \arrow[d, dotted] \\
    \framep[n][\lbrack 0,p-1 \rbrack] &
    \restrf[n][\lbrack 0,p-1 \rbrack] \arrow[l, dotted] \arrow[uul, dotted, "\mbox{$[0,p-1]$}"', near start, shift right=2mm, hook']\\
  \end{tikzcd}
\end{center}

In particular, the dependency of $\framep[n+1][\lbrack 0,p-1 \rbrack]$ within the type of $\restrf[n][\lbrack 0,p-1 \rbrack]$ combined with the dependency of $\restrf[n][\lbrack 0,p-1 \rbrack]$ within the definition of $\framep[n+1][\lbrack 0,p \rbrack]$ requires a mutual definition of the type of $\restrf[n][\lbrack 0,p-1 \rbrack]$ and of $\framep[n+1][p]$ as a function from $\restrf[n][\lbrack 0,p-1 \rbrack]$.

Using this recursive definition of $\framep[n+1][\lbrack 0,p \rbrack]$ and $\restrf[n][\lbrack 0,p-1 \rbrack]$, we are thus able to extend the specification of the structure with its third component, for all $p \leq n$.

\begin{equation*}
  \smfontsize
  \begin{array}{lcl}
    \framep[n][p]   & : & \HSet                                             \\
    \painting[n][p] & : & \framep[n][p] \rightarrow \HSet                   \\
    \restrf[n][p]   & : & \Pi q \leq n-p.\, \Pi \epsilon.\, \framep[n+1][p]
    \left(\begin{array}{l}
              \framep[n][\lbrack 0,p-1 \rbrack]   \\
              \painting[n][\lbrack 0,p-1 \rbrack] \\
              \restrf[n][\lbrack 0,p-1 \rbrack]   \\
            \end{array}\right)
    \rightarrow \framep[n][p]                                               \\
  \end{array}
\end{equation*}
This can be reformulated using sequences for appropriate definitions of $\frametype[n][\lbrack 0,p \rbrack]$, $\paintingtype[n][\lbrack 0,p \rbrack]$, and $\restrftype[n][\lbrack 0,p \rbrack]$:
\begin{equation*}
  \smfontsize
  \begin{array}{lcl}
    \framep[n][\lbrack 0,p \rbrack]   & : & \frametype[n][\lbrack 0,p \rbrack]                                     \\
    \painting[n][\lbrack 0,p \rbrack] & : & \paintingtype[n][\lbrack 0,p \rbrack](\framep[n][\lbrack 0,p \rbrack]) \\
    \restrf[n][\lbrack 0,p \rbrack]   & : & \restrftype[n][\lbrack 0,p \rbrack]
    \left(\begin{array}{l}
              \framep[n][\lbrack 0,p-1 \rbrack]   \\
              \painting[n][\lbrack 0,p-1 \rbrack] \\
              \restrf[n][\lbrack 0,p-1 \rbrack]   \\
            \end{array}\right)                                                                      \\
  \end{array}
\end{equation*}

Types such as $\restrftype[n][\lbrack 0,p \rbrack]$ will be represented by iterated $\Sigma$-types since each $\restrf[n][p]$ may depend on the previous components of the sequence. In the case of $\framep$ and $\painting$, there is no dependency, so an ordinary product can be used to represent $\frametype[n][\lbrack 0,p \rbrack]$ and $\paintingtype[n][\lbrack 0,p \rbrack]$.

Let us turn now to the specification of $\restrp[n][p]$. Since it depends on $\painting[n+1][p]$, we need first to define $\painting[n+1][p]$ but in addition to the type dependency in $\framep[n+1][p]$, this requires the definition of $\painting[n][p-1]$, $\framep[n][p-1]$ by typing dependency, and $\restrf[n][p-1]$. Since $\painting[n+1][p]$ also depends on $\painting[n+1][p+1]$ when $p < n+1$, these dependencies are actually between $\painting[n+1][\lbrack p,n+1 \rbrack]$, $\painting[n][\lbrack p,n \rbrack]$, $\framep[n+1][\lbrack p,n+1 \rbrack]$, $\framep[n][\lbrack p,n \rbrack]$, and $\restrf[n][\lbrack p,n \rbrack]$. For $p = n$, this requires the assumption of an inhabitant $E_{n+1}$ of $\Xcomp{m}{n+1}$. Finally, $\framep[n+1][\lbrack p,n+1 \rbrack]$ depends on $\framep[n+1][\lbrack 0,p-1 \rbrack]$, $\framep[n+1][\lbrack p,n \rbrack]$ on $\framep[n][\lbrack 0,p-1 \rbrack]$, and $\restrf[n][\lbrack p,n \rbrack]$ on $\restrf[n][\lbrack 0,p-1 \rbrack]$, leading to the following graph of dependencies where indirect dependencies in definitions are now dashed lines.

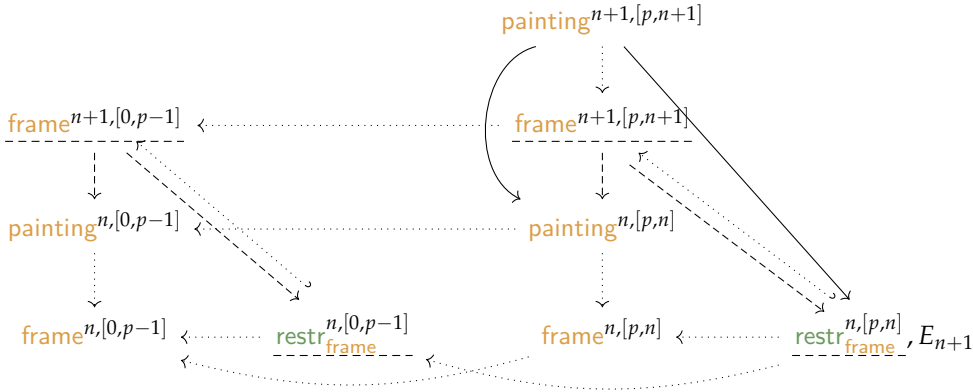
\begin{center}
  \begin{tikzcd}
    & & \painting[n+1][\lbrack p,n+1 \rbrack] \arrow[dd, bend right=70] \arrow[dddr] \arrow[d, dotted]\\
    \udensdash{\framep[n+1][\lbrack 0,p-1 \rbrack]} \arrow[d, dashed] \arrow[ddr, dashed] & &
    \udensdash{\framep[n+1][\lbrack p,n+1 \rbrack]} \arrow[ll, dotted] \arrow[d, dashed] \arrow[ddr, dashed, shift right=2mm] \\
    \painting[n][\lbrack 0,p-1 \rbrack] \arrow[d, dotted] & &
    \painting[n][\lbrack p,n \rbrack] \arrow[ll, dotted] \arrow[d, dotted] \\
    \framep[n][\lbrack 0,p-1 \rbrack] &
    \udensdash{\restrf[n][\lbrack 0,p-1 \rbrack]} \arrow[l, dotted] \arrow[uul, dotted, shift right=2mm, hook'] &
    \framep[n][\lbrack p,n \rbrack] \arrow[ll, dotted, bend left=15] &
    \udensdash{\restrf[n][\lbrack p,n \rbrack]},E_{n+1}\arrow[ll, dotted, bend left=15] \arrow[l, dotted] \arrow[uul, dotted, hook'] \\
  \end{tikzcd}
\end{center}

This is eventually enough to specify the type $\restrptype[n][\lbrack 0,p \rbrack]$ of $\restrp[n][\lbrack 0,p \rbrack]$ so that we can focus on the specification of $\cohf[n][p]$. An analysis similar to the one done for $\restrf[n][p]$ leads to conclude that the specification of $\cohf[n][\lbrack 0,p \rbrack]$ should be done mutually with the definition of $\restrf[n+1][\lbrack 0,p+1 \rbrack]$.

Then, $\cohp[n][p]$ has to be specified, for which the same analysis done for specifying $\restrp[n][p]$ applies. In particular, it requires first to define $\restrp[n+1][\lbrack 0,p \rbrack]$.

The specification of the six fields being complete, the induction step can be built. All of $\framep[n+1][\lbrack 0,p \rbrack]$, $\painting[n+1][\lbrack 0,p \rbrack]$, $\restrf[n+1][\lbrack 0,p \rbrack]$ and $\restrp[n+1][\lbrack 0,p \rbrack]$ have already been built, so only the construction of $\cohf[n+1][\lbrack 0,p \rbrack]$ and $\cohp[n+1][\lbrack 0,p \rbrack]$ remain. Details follow.

\subsection{Specifying frames\href{https://artagnon.github.io/bonak/docs/Bonak.\%CE\%BDSet.\%CE\%BDSet.html\#mkFrameTypes}{\linkicon}}
Assuming $\framep[n][\lbrack 0,p-1 \rbrack]$ for $p \leq n+1$ only requires defining the type $\frametype[n][\lbrack 0,p-1 \rbrack]$ of frames, which is just a $p$-ary product of $\HSet$.

\begin{equation*}
  \smfontsize
  \begin{array}{llcl}
    \frametype[n][\lbrack 0,p-1 \rbrack] &  & :      & \HSet                                             \\
    \frametype[n]                        &  & \defeq & \unittype                                         \\
    \frametype[n][\lbrack 0,p \rbrack]   &  & \defeq & \frametype[n][\lbrack 0,p-1 \rbrack] \times \HSet \\
  \end{array}
\end{equation*}

\subsection{Specifying paintings\href{https://artagnon.github.io/bonak/docs/Bonak.\%CE\%BDSet.\%CE\%BDSet.html\#mkPaintingTypes}{\linkicon}}

The type $\paintingtype[n][\lbrack 0,p-1 \rbrack]$ of $\painting[n][\lbrack 0,p-1 \rbrack]$ for $p \leq n+1$ is a product of type families over frames, so $\framep[n][\lbrack 0,p-1 \rbrack]$ needs to be assumed.

\begin{equation*}
  \smfontsize
  \begin{array}{llcl}
    \paintingtype[n][\lbrack 0,p-1 \rbrack] & \framep[n][\lbrack 0,p-1 \rbrack] & :      & \HSet                                                                                                               \\
    \paintingtype[n]                        & \unitpoint                        & \defeq & \unittype                                                                                                           \\
    \paintingtype[n][\lbrack 0,p \rbrack]   & \framep[n][\lbrack 0,p \rbrack]   & \defeq & \paintingtype[n][\lbrack 0,p-1 \rbrack](\framep[n][\lbrack 0,p-1 \rbrack]) \times (\framep[n][p] \rightarrow \HSet) \\
  \end{array}
\end{equation*}

\subsection{Defining frames and specifying frame restrictions\href{https://artagnon.github.io/bonak/docs/Bonak.\%CE\%BDSet.\%CE\%BDSet.html\#mkRestrFrameTypesAndFrames}{\linkicon}\label{sec:mkframe}}

Assuming $\framep[n][\lbrack 0,p-1 \rbrack]$ and $\painting[n][\lbrack 0,p-1 \rbrack]$ for $p \leq n+1$, we can mutually define the types $\restrftype[n][\lbrack 0,p-1 \rbrack]$ together with $\framep[n+1][\lbrack 0,p \rbrack]$, as a function from $\restrf[n][\lbrack 0,p-1 \rbrack]$. We introduce the abbreviation $\depstype[n][\lbrack 0,p-1 \rbrack]$ for the combination of sequences of $\framep[n][\lbrack 0,p-1 \rbrack]$, $\painting[n][\lbrack 0,p-1 \rbrack]$ and $\restrf[n][\lbrack 0,p-1 \rbrack]$.

Once a $\deps: \depstype$ is assumed, we freely refer to $\framep$, $\painting$, $\restrf$.

\begin{equation*}
  \smfontsize
  \begin{array}{llcl}
    \depstype[n][\lbrack 0,p-1 \rbrack]    &        & \defeq                               &
    \left(\begin{array}{lll}
              \framep[n][\lbrack 0,p-1 \rbrack]   & : & \frametype[n][\lbrack 0,p-1 \rbrack]    \\
              \painting[n][\lbrack 0,p-1 \rbrack] & : & \paintingtype[n][\lbrack 0,p-1 \rbrack]
              (\framep[n][\lbrack 0,p-1 \rbrack])                                               \\
              \restrf[n][\lbrack 0,p-1 \rbrack]   & : & \restrftype[n][\lbrack 0,p-1 \rbrack]
              \left(\begin{array}{l}
                  \framep[n][\lbrack 0,p-1 \rbrack]   \\
                  \painting[n][\lbrack 0,p-1 \rbrack] \\
                \end{array}\right)                                         \\
            \end{array}\right)  \\
    \framep[n+1][\lbrack 0,p \rbrack]      &
    \deps[n][\lbrack 0,p-1 \rbrack]        & :      & \frametype[n+1][\lbrack 0,p \rbrack]   \\
    \framep[n+1][\lbrack 0,0 \rbrack]      &
    \deps[n]                               & \defeq & (\unitpoint,\unittype)                 \\
    \framep[n+1][\lbrack 0,p+1 \rbrack]    &
    \deps[n][\lbrack 0,p \rbrack]          & \defeq &
    \left(\begin{array}{l}
              \framep[n+1][\lbrack 0,p \rbrack]
              (\deps[n][\lbrack 0,p-1 \rbrack]), \\
              \Sigma d:\framep[n+1][p]
              (\deps[n][\lbrack 0,p-1 \rbrack]).
              \, \layer[n][p]
              (\deps[n][\lbrack 0,p \rbrack])
              (d)                                \\
            \end{array}\right)                                                 \\
    \restrftype[n][\lbrack 0,p-1 \rbrack]  &
    \left(\begin{array}{l}
              \framep[n][\lbrack 0,p-1 \rbrack]   \\
              \painting[n][\lbrack 0,p-1 \rbrack] \\
            \end{array}\right) & :      & \HSet                                                \\
    \restrftype[n]                         &
    \left(\begin{array}{l}
              \unitpoint \\
              \unitpoint \\
            \end{array}\right)                 & \defeq &
    \unittype                                                                                \\
    \restrftype[n][\lbrack 0,p \rbrack]    &
    \left(\begin{array}{l}
              \framep[n][\lbrack 0,p \rbrack]   \\
              \painting[n][\lbrack 0,p \rbrack] \\
            \end{array}\right)   & \defeq &
    \left\{\begin{array}{l}
             \Sigma \restrf[n][\lbrack 0,p-1 \rbrack]
             :\restrftype[n][\lbrack 0,p-1 \rbrack]
             \left(\begin{array}{l}
                \framep[n][\lbrack 0,p-1 \rbrack]   \\
                \painting[n][\lbrack 0,p-1 \rbrack] \\
              \end{array}\right).         \\
             \Pi q \leq n-p.\,\Pi \epsilon.                      \\
             \framep[n+1][p]
             \left(\begin{array}{l}
                \framep[n][\lbrack 0,p-1 \rbrack]   \\
                \painting[n][\lbrack 0,p-1 \rbrack] \\
                \restrf[n+1][\lbrack 0,p-1 \rbrack] \\
              \end{array}\right) \rightarrow  \framep[n][p] \\
           \end{array}\right.                \\
    \layertype[n][p]                       &
    \deps[n][\lbrack 0,p \rbrack]
                                           & :      & \HSet                                  \\
    \layertype[n][p]                       &
    \deps[n][\lbrack 0,p \rbrack]
                                           & \defeq & \framep[n+1][p]
    (\deps[n][\lbrack 0,p-1 \rbrack])
    \rightarrow \HSet                                                                        \\
    \layer[n][p]                           &
    \deps[n][\lbrack 0,p \rbrack]
                                           & :      & \layertype[n][p]
    (\deps[n][\lbrack 0,p \rbrack])
    \\
    \layer[n][p]                           &
    \deps[n][\lbrack 0,p \rbrack]
                                           & \defeq &
    \lambda d.\, \Pi\epsilon.\,\painting[n][p](\restrf[n][p][0][\epsilon](d))                \\
  \end{array}
\end{equation*}

\subsection{Defining paintings\href{https://artagnon.github.io/bonak/docs/Bonak.\%CE\%BDSet.\%CE\%BDSet.html\#mkPaintings}{\linkicon}\label{sec:mkpainting}}

To define $\painting[n+1][p]$, let us introduce another abbreviation that combines a full $\deps[n][\lbrack 0,n \rbrack]:\depstype[n][\lbrack 0,n \rbrack]$ with a family $E:\framep[n+1][n+1](\deps[n][\lbrack 0,n \rbrack])\rightarrow \HSet$.

\begin{equation*}
  \smfontsize
  \fulldepstype[n] \defeq
  \left(\begin{array}{lcl}
      \deps[n][\lbrack 0,n \rbrack] & : & \depstype[n][\lbrack 0,n \rbrack]                                  \\
      E                             & : & \framep[n+1][n+1](\deps[n][\lbrack 0,n \rbrack]) \rightarrow \HSet \\
    \end{array}\right)
\end{equation*}

Then, $\painting[n+1][p]$ is defined for $p \leq n+1$ by recursion from the base case $n+1$, and down to $p$.

\begin{equation*}
  \smfontsize
  \begin{array}{llcl}
    \painting[n+1][p]        &
    \fulldeps[n]             & :      & \paintingtype[n+1][p] \\
    \painting[n+1][n+1]      &
    \fulldeps[n]             & \defeq & E                     \\
    \painting[n+1][p \leq n] &
    \fulldeps[n]             & \defeq & \lambda d.\,
    \Sigma l:\layer[n][p](\deps[n][\lbrack 0,p \rbrack]).
    \painting[n+1][p+1]
    (\fulldeps[n])(d,l)                                       \\
  \end{array}
\end{equation*}

From this, we can deduce $\painting[n+1][\lbrack 0,p-1 \rbrack]$ for $p \leq n+2$ by a second recursion, this time on $p$.

\begin{equation*}
  \smfontsize
  \begin{array}{llcl}
    \painting[n+1][\lbrack 0,p-1 \rbrack] &
    \fulldeps[n]                          & :      & \paintingtype[n+1][\lbrack 0,p-1 \rbrack](\framep[n+1][\lbrack 0,p-1 \rbrack](\deps[n][\lbrack 0,p-1 \rbrack])) \\
    \painting[n+1]                        &
    \fulldeps[n]                          & \defeq & \unittype                                                                                                       \\
    \painting[n+1][\lbrack 0,p \rbrack]   &
    \fulldeps[n]
                                          & \defeq &
    (\painting[n+1][\lbrack 0,p-1 \rbrack](\fulldeps[n]),\painting[n+1][p](\fulldeps[n]))                                                                            \\
  \end{array}
\end{equation*}

\subsection{Specifying painting restrictions\href{https://artagnon.github.io/bonak/docs/Bonak.\%CE\%BDSet.\%CE\%BDSet.html\#mkRestrPaintingTypes}{\linkicon}}

To go further, we need to specify the type of sequences of painting restrictions, for $p \leq n+1$.

\begin{equation*}
  \smfontsize
  \begin{array}{llcl}
    \restrptype[n][\lbrack 0,p-1 \rbrack] &
    \fulldeps[n]                          & :            & \HSet                                                    \\
    \restrptype[n]                        & \fulldeps[n] & \defeq                                       & \unittype \\
    \restrptype[n][\lbrack 0,p \rbrack]   &
    \fulldeps[n]
                                          & \defeq       & \restrptype[n][\lbrack 0,p-1 \rbrack] \times
    \left(
    \begin{array}{l}
        \Pi q \leq n-p.\Pi \epsilon.\Pi d:\framep[n+1][p](\deps[n][\lbrack 0,p-1 \rbrack]). \\
        \painting[n+1][p]
        (\fulldeps[n])(d) \rightarrow                                                       \\ \painting[n][p](\restrf[n][p][q][\epsilon](d))\end{array}
    \right)                                                                                                         \\
  \end{array}
\end{equation*}

\subsection{Defining frame restrictions and specifying frame coherence laws\href{https://artagnon.github.io/bonak/docs/Bonak.\%CE\%BDSet.\%CE\%BDSet.html\#mkCohFrameTypesAndRestrFrames}{\linkicon}\label{sec:mkrestrf}}

We can now mutually define $\restrf[n+1][\lbrack 0,p \rbrack]$ and specify $\cohftype[n][\lbrack 0,p-1 \rbrack]$ from $\framep[n][\lbrack 0,n \rbrack]$, $\painting[n][\lbrack 0,n \rbrack]$, $\restrf[n][\lbrack 0,n \rbrack]$, $\restrp[n][\lbrack 0,p-1 \rbrack]$ and some $E:\framep[n+1][n+1](\deps[n][\lbrack 0,n \rbrack]) \rightarrow \HSet$, abbreviating these dependencies and the dependency in $\cohf[n][\lbrack 0,p-1 \rbrack]$ as $\depscohstype[n][\lbrack 0,p-1 \rbrack]$, and using abbreviations in passing for $\restrl[n][p]$ and $\restrltype[n][p]$. It is convenient to build $\deps[n+1][\lbrack 0,p \rbrack]$ from $\depscohs[n][\lbrack 0,p-1 \rbrack]$, so that we can now build $\framep[n+2][p]$ one step further.

\begin{equation*}
  \smfontsize
  \begin{array}{llcl}
    \depscohstype[n][\lbrack 0,p-1 \rbrack] &                                     & \defeq                                                         &
    \left(\begin{array}{lcl}
              \fulldeps[n]                      & : & \fulldepstype[n]                                    \\
              \restrp[n][\lbrack 0,p-1 \rbrack] & : & \restrptype[n][\lbrack 0,p-1 \rbrack](\fulldeps[n]) \\
              \cohf[n][\lbrack 0,p-1 \rbrack]   & : & \cohftype[n][\lbrack 0,p-1 \rbrack]
              \left(\begin{array}{l}
                  \fulldeps[n]                      \\
                  \restrp[n][\lbrack 0,p-1 \rbrack] \\
                \end{array}\right)                                                     \\
            \end{array}\right)                                                \\
    \overline{\deps}                        &
    \depscohs[n][\lbrack 0,p-1 \rbrack]     & :                                   & \depstype[n+1][\lbrack 0,p \rbrack]                              \\
    \overline{\deps}                        & \depscohs[n][\lbrack 0,p-1 \rbrack] & \defeq                                                         &
    \left(\begin{array}{l}
              \framep[n+1][\lbrack 0,p \rbrack](\deps[n][\lbrack 0,p-1 \rbrack])     \\
              \painting[n+1][\lbrack 0,p \rbrack](\fulldeps[n])                      \\
              \restrf[n+1][\lbrack 0,p \rbrack](\depscohs[n][\lbrack 0,p-1 \rbrack]) \\
            \end{array}\right)                                                                     \\
    \restrf[n+1][\lbrack 0,p \rbrack]       &
    \depscohs[n][\lbrack 0,p-1 \rbrack]     & :                                   &
    \restrftype[n+1][\lbrack 0,p \rbrack]
    \left(\begin{array}{l}
              \framep[n+1][\lbrack 0,p \rbrack](\deps[n][\lbrack 0,p \rbrack]) \\
              \painting[n+1][\lbrack 0,p \rbrack](\fulldeps[n])                \\
            \end{array}\right)                                                                           \\
    \restrf[n+1][\lbrack 0,0 \rbrack]       &
    \depscohs[n]    
                                            & \defeq                              & (\unitpoint,\lambda q. \lambda\epsilon. \lambda \_.\unitpoint)   \\
    \restrf[n+1][\lbrack 0,p+1 \rbrack]     &
    \depscohs[n][\lbrack 0,p \rbrack]       & \defeq                              &
    \left(\begin{array}{l}
              \restrf[n+1][\lbrack 0,p \rbrack][q][\epsilon]
              (\depscohs[n][\lbrack 0,p-1 \rbrack]),
              \lambda q.\lambda \epsilon.\lambda (d,l).        \\
              \left(\restrf[n+1][p][q][\epsilon]
                (\depscohs[n][\lbrack 0,p-1 \rbrack])(d), \restrl[n][p][q][\epsilon]
              (\depscohs[n][\lbrack 0,p \rbrack])(d)(l)\right) \\
            \end{array} \right)                                                                              \\
    \cohftype[n][\lbrack 0,p-1 \rbrack]     &
    \left(\begin{array}{l}
              \fulldeps[n]                      \\
              \restrp[n][\lbrack 0,p-1 \rbrack] \\
            \end{array}\right)    & :                                   & \HSet                                                                        \\
    \cohftype[n]                            &
    \left(\begin{array}{l}
              \fulldeps[n] \\
              \unitpoint   \\
            \end{array}\right)                  & \defeq                              &
    \unittype                                                                                                                                        \\
    \cohftype[n][\lbrack 0,p \rbrack]       &
    \left(\begin{array}{l}
              \fulldeps[n]                    \\
              \restrp[n][\lbrack 0,p \rbrack] \\
            \end{array}\right)      & \defeq                              &
    \left\{\begin{array}{l}
             \Sigma \cohf[n][\lbrack 0,p-1 \rbrack][q][r][\epsilon][\omega]
             :\cohftype[n][\lbrack 0,p-1 \rbrack]
             \left(\begin{array}{l}
                \fulldeps[n]                      \\
                \restrp[n][\lbrack 0,p-1 \rbrack] \\
              \end{array}\right).                          \\
             \Pi d:\framep[n+2][p]
             \left(\overline{\deps}\left(\begin{array}{l}
                                        \fulldeps[n]                      \\
                                        \restrp[n][\lbrack 0,p-1 \rbrack] \\
                                        \cohf[n][\lbrack 0,p-1 \rbrack]
                                      \end{array}\right)\right)          . \\
             \\
             \restrf[n][p][q][\epsilon](\restrf[n+1][p][r][\omega]
             \left(\begin{array}{l}
                \fulldeps[n]                      \\
                \restrp[n][\lbrack 0,p-1 \rbrack] \\
                \cohf[n][\lbrack 0,p-1 \rbrack]
              \end{array}\right)(d)) =                          \\
             \restrf[n][p][r][\omega](\restrf[n+1][p][q+1][\epsilon]
             \left(\begin{array}{l}
                \fulldeps[n]                      \\
                \restrp[n][\lbrack 0,p-1 \rbrack] \\
                \cohf[n][\lbrack 0,p-1 \rbrack]
              \end{array}\right)(d))                          \\
           \end{array}\right.                                                                    \\
    \restrltype[n][p]                       &
    \depscohs[n][\lbrack 0,p \rbrack]
                                            & :                                   & \HSet                                                            \\
    \restrltype[n][p]                       &
    \depscohs[n][\lbrack 0,p \rbrack]
                                            & \defeq                              &
    \left\{\begin{array}{l}
             \Pi q \leq n-p. \Pi \epsilon. \Pi d: \framep[n+2][p](\overline{\deps}(\depscohs[n][\lbrack 0,p-1 \rbrack])).\, \\
             \layer[n+1][p](\overline{\deps}(\depscohs[n][\lbrack 0,p-2 \rbrack])(d) \rightarrow                            \\
             \layer[n][p](\deps[n][\lbrack 0,p \rbrack])(\restrf[n][p]
             (\depscohs[n][\lbrack 0,p-1 \rbrack])(d))
           \end{array}\right.                            \\
    \restrl[n][p]                           &
    \depscohs[n][\lbrack 0,p \rbrack]       & :                                   &
    \restrltype[n][p] (\depscohs[n][\lbrack 0,p \rbrack])                                                                                            \\
    \restrl[n][p][q][\epsilon]              &
    \depscohs[n][\lbrack 0,p \rbrack]       & \defeq                              &
    \begin{array}{l}
      \lambda d.\lambda l.\lambda \omega.\overrightarrow{\cohf[n][p][q][0][\epsilon][\omega](d)}               \\
      \restrp[n][p][q][\epsilon](\restrf[n][p][q][\epsilon](\depscohs[n][\lbrack 0,p-1 \rbrack])(d))(l_\omega) \\
    \end{array}
  \end{array}
\end{equation*}

\subsection{Defining painting restrictions\href{https://artagnon.github.io/bonak/docs/Bonak.\%CE\%BDSet.\%CE\%BDSet.html\#mkRestrPaintings}{\linkicon}\label{sec:mkrestrp}}

Armed with $\restrf[n+1][\lbrack 0,p \rbrack]$ and $\cohftype[n][\lbrack 0,p-1 \rbrack]$, thus also with $\restrl[n][\lbrack 0,p-1 \rbrack]$, it is possible to define $\restrp[n+1][p]$. First, we define a notion of full dependencies at the level of coherence conditions, that combines a full $\depscohs[n][\lbrack 0,n \rbrack]$ with a set $E':\framep[n+2][n+2](\overline{\deps}(\depscohs[n][\lbrack 0,n \rbrack])) \rightarrow \HSet$.

\begin{equation*}
  \smfontsize
  \fulldepscohstype[n] \defeq
  \left(\begin{array}{lcl}
      \depscohs[n][\lbrack 0,n \rbrack] & : & \depscohstype[n][\lbrack 0,n \rbrack]                                                   \\
      E'                                & : & \framep[n+2][n+2](\overline{\deps}(\depscohs[n][\lbrack 0,n \rbrack]))\rightarrow \HSet \\
    \end{array}\right)
\end{equation*}

We can then extend $\overline{\deps}$ to build $\fulldeps$ from $\fulldepscohs$.

\begin{equation*}
  \smfontsize
  \begin{array}{llcl}
    \overline{\fulldeps} & \fulldepscohs[n] & :      & \fulldepstype[n+1] \\
    \overline{\fulldeps} & \fulldepscohs[n] & \defeq & \left(
    \begin{array}{l}
        \overline{\deps}(\depscohs[n][\lbrack 0,n \rbrack]) \\
        E'                                                  \\
      \end{array}\right)                   \\
  \end{array}
\end{equation*}

Then we can define $\restrp[n+1][p]$ for $p \leq n+1$ by recursion on $n+1-p$.

\begin{equation*}
  \smfontsize
  \begin{array}{llcl}
    \restrp[n+1][p][q][\epsilon]        &
    \fulldepscohs[n]                    & :      &
    \left(
    \begin{array}{l}
        \Pi d:\framep[n+2][p]
        (\overline{\deps}(\depscohs[n][\lbrack 0,p-2 \rbrack])). \\
        \painting[n+2][p]
        (\overline{\fulldeps}(\fulldepscohs[n]))(d) \rightarrow  \\
        \painting[n+1][p](\fulldeps[n])
        (\restrf[n+1][p](\depscohs[n][\lbrack 0,p-1 \rbrack])(d))\end{array}\right)                                         \\
    \restrp[n+1][n+1][q][\epsilon]      &
    \fulldepscohs[n]                    & \defeq & \lambda d. \lambda (l,c). l_\epsilon \\
    \restrp[n+1][p \leq n][q][\epsilon] &
    \fulldepscohs[n]                    & \defeq & \lambda d. \lambda (l,c).
    \left(\begin{array}{l}
              \restrl[n][p][q][\epsilon](\depscohs[n][\lbrack 0,p \rbrack])(d)(l), \\
              \restrp[n+1][p+1][q][\epsilon](\fulldepscohs[n])(d,l)(c)             \\
            \end{array}\right)          \\
  \end{array}
\end{equation*}

From this, we can deduce $\restrp[n+1][\lbrack 0,p-1 \rbrack]$ for $p \leq n+2$ again by a second recursion, this time on $p$.

\begin{equation*}
  \smfontsize
  \begin{array}{llcl}
    \restrp[n+1][\lbrack 0,p-1 \rbrack] &
    \fulldepscohs[n]                    & :      & \restrptype[n+1][\lbrack 0,p-1 \rbrack](\fulldeps[n]) \\
    \restrp[n+1]                        &
    \fulldepscohs[n]                    & \defeq & \unittype                                             \\
    \restrp[n+1][\lbrack 0,p \rbrack]   &
    \fulldepscohs[n]
                                        & \defeq &
    (\restrp[n+1][\lbrack 0,p-1 \rbrack](\fulldepscohs[n]),\restrp[n+1][p](\fulldepscohs[n]))            \\
  \end{array}
\end{equation*}

\subsection{Specifying painting coherence laws\href{https://artagnon.github.io/bonak/docs/Bonak.\%CE\%BDSet.\%CE\%BDSet.html\#mkCohPaintingTypes}{\linkicon}}

To go further, we need to specify the type $\cohptype[n][\lbrack 0,p-1 \rbrack]$ of painting coherence laws, for $p \leq n+1$.

\begin{equation*}
  \smfontsize
  \begin{array}{llcl}
    \cohptype[n][\lbrack 0,p-1 \rbrack] & \fulldepscohs[n] & :      & \HSet                                                        \\
    \cohptype[n]                        & \unitpoint       & \defeq & \unittype                                                    \\
    \cohptype[n][\lbrack 0,p \rbrack]   & \fulldepscohs[n] & \defeq & \cohptype[n][\lbrack 0,p-1 \rbrack](\fulldepscohs[n]) \times \\
    \multicolumn{4}{c}{\left\{
      \begin{array}{l}
        \Pi q \leq n-p.\,\Pi r\leq q.\,\Pi \epsilon.\,\Pi \omega.                                                                                       \\
        \Pi d:\framep[n+2][p](\overline{\deps}(\depscohs[n][\lbrack 0,p-2 \rbrack])).\,\Pi c:\painting[n+2][p](\overline{\fulldeps}(\fulldepscohs[n])). \\
        \restrp[n][p][q][\epsilon](\restrf[n+1][p](\depscohs[n][\lbrack 0,p-1 \rbrack])(d))(\restrp[n+1][p][r][\omega]
        (\fulldepscohs[n])(d)(c)) =_{\cohf[n][p][q][r][\epsilon][\omega]}                                                                               \\
        \restrp[n][p][r][\omega](\restrf[n+1][p](\depscohs[n][\lbrack 0,p-1 \rbrack])(d))(\restrp[n+1][p][q+1][\epsilon]
        (\fulldepscohs[n])(d)(c))
      \end{array}
    \right.}                                                                                                                       \\
  \end{array}
\end{equation*}
where we use the ``equality over'' notation $t =_e u$ to mean $\overrightarrow{e}(t) = u$ whenever $t$ is of some type $P(a)$, $u$ of type $P(b)$ and $e$ a proof of $a = b$.

\subsection{Defining frame coherence laws and specifying frame 2-dimensional coherence laws\href{https://artagnon.github.io/bonak/docs/Bonak.\%CE\%BDSet.\%CE\%BDSet.html\#mkCohFrames}{\linkicon}\label{sec:mkcohf}}

With $\cohptype[n][\lbrack 0,p-1 \rbrack]$, we can mutually define $\cohf[n+1][\lbrack 0,p \rbrack]$ and $\cohttype[n][\lbrack 0,p-1 \rbrack]$ for $p \leq n+1$ from $\fulldepscohs[n]$. Due to reasoning in $\HSet$, we do not need to have $\cohf[n+1][\lbrack 0,p \rbrack]$ depending on a proof of $\cohttype[n][\lbrack 0,p-1 \rbrack]$ as the latter can be proved by uniqueness of identity proofs. It is enough to consider $\cohttype[n][p]$ instead of $\cohttype[n][\lbrack 0,p-1 \rbrack]$.

It is also convenient to build $\depscohs[n+1][\lbrack 0,p \rbrack]$ from $\depscoht[n][\lbrack 0,p-1 \rbrack]$, so that we can now build $\framep[n+3][p]$. We implicitly use in $\cohf[n+1][\lbrack 0,p \rbrack]$ that equalities in a $\Sigma$-type are equivalent to the $\Sigma$-type of the underlying equalities.

\begin{equation*}
  \smfontsize
  \begin{array}{llcl}
    \depscohttype[n][\lbrack 0,p-1 \rbrack]     &                                     & \defeq                                                                            &
    \left(\begin{array}{lll}
              \fulldepscohs[n]                & : & \fulldepscohstype[n]                                  \\
              \cohp[n][\lbrack 0,p-1 \rbrack] & : & \cohptype[n][\lbrack 0,p-1 \rbrack](\fulldepscohs[n]) \\
            \end{array}
    \right)                                                                                                                                                                                                         \\
    \overline{\depscohs}                        & \depscoht[n][\lbrack 0,p-1 \rbrack] & :                                                                                 & \depscohstype[n+1][\lbrack 0,p \rbrack] \\
    \overline{\depscohs}                        & \depscoht[n][\lbrack 0,p-1 \rbrack] & \defeq                                                                            &
    \left(\begin{array}{l}
              \overline{\fulldeps}(\fulldepscohs[n])             \\
              \restrp[n+1][p](\fulldepscohs[n])                  \\
              \cohf[n+1][p](\depscoht[n][\lbrack 0,p-1 \rbrack]) \\
            \end{array}\right)                                                                                                                                                        \\
    \cohf[n+1][\lbrack 0,p \rbrack]             &
    \depscoht[n][\lbrack 0,p-1 \rbrack]         & :                                   & \cohftype[n+1][\lbrack 0,p \rbrack]
    \left(\begin{array}{l}
              \overline{\fulldeps}(\fulldepscohs[n])              \\
              \restrp[n+1][\lbrack 0,p \rbrack](\fulldepscohs[n]) \\
            \end{array}\right)                                                                                                                                                       \\
    \\
    \cohf[n+1][\lbrack 0,0 \rbrack]             &
    \depscoht[n]                                & \defeq                              & (\unitpoint,\lambda q.\lambda r. \lambda\epsilon. \lambda\omega.\lambda \_.\refl)                                           \\
    \cohf[n+1][\lbrack 0,p+1 \rbrack]           &
    \depscoht[n][\lbrack 0,p \rbrack]           & \defeq                              &
    \left(\begin{array}{l}
              \cohf[n+1][\lbrack 0,p \rbrack]
              (\depscoht[n][\lbrack 0,p-1 \rbrack]),                             \\
              \lambda q.\lambda r.\lambda \epsilon.\lambda \omega.\lambda (d,l). \\
              (\cohf[n+1][p][q][r][\epsilon][\omega]
              (\depscoht[n][\lbrack 0,p \rbrack])(d), \cohl[n][p][q][r][\epsilon][\omega]
              (\depscoht[n][\lbrack 0,p \rbrack])(d)(l))                         \\
            \end{array}\right)                                                                                                                                \\
    \coht[n][p]                                 &
    \fulldepscohs[n]                            & :                                   & \cohttype[n][p](\fulldepscohs[n])                                                                                           \\
    \coht[n][p][q][r][\epsilon][\omega][\theta] &
    \fulldepscohs[n]                            & \defeq                              & \lambda d.\,\UIP                                                                                                            \\

    \cohl[n][p]                                 &
    \depscoht[n][\lbrack 0,p \rbrack]           & :                                   &
    \left\{\begin{array}{l}
             \Pi q \leq n-p.\,\Pi r\leq q.\,\Pi \epsilon.\,\Pi \omega.\,\Pi d.\,\Pi l.                                                                               \\
             \restrl[n][p][q][\epsilon](\depscohs[n][\lbrack 0,p \rbrack])(\restrl[n+1][p][r][\omega](\overline{\depscohs}(\depscoht[n][\lbrack 0,p-2 \rbrack]))(l)) \\ =_{\cohf[n+1][p][r+1][q+1][\epsilon][\omega](\depscoht[n][\lbrack 0,p-1 \rbrack])(d)} \\
             \restrl[n][p][r][\omega](\depscohs[n][\lbrack 0,p \rbrack])(\restrl[n+1][p][q+1][\epsilon](\overline{\depscohs}(\depscoht[n][\lbrack 0,p-2 \rbrack]))(l))
           \end{array}\right.                                    \\
    \cohl[n][p][q][r][\epsilon][\omega]         &
    \depscoht[n][\lbrack 0,p \rbrack]           & \defeq                              &
    \left\{\begin{array}{l}
             \lambda d. \lambda l. \lambda \theta. \overrightarrow{\coht[n][p][q][r][\epsilon][\omega][\theta](\fulldepscohs[n])(d)}                             \\
             (\ap \overrightarrow{\cohf[n][p][r][0][\omega][\theta](\restrf[n+2][p][q+2][\epsilon](\overline{\depscohs}(\depscoht[n][\lbrack 0,p \rbrack]))(d))} \\ (\ap \overrightarrow{\ap\restrf[n][p][r][\omega](\cohf[n+1][p][q+1][0][\epsilon][\theta](\depscoht[n][\lbrack 0,p-1 \rbrack])(d))} \\(\cohp[n][p][q][r][\epsilon][\omega](l_\theta)))) \\
           \end{array}\right.
  \end{array}
\end{equation*}
where $\ap$ applies a function on both sides of an equality and $\cohttype[n][p]$ is:
\begin{equation*}
  \smfontsize
  \begin{array}{llcl}
    \cohttype[n][p] & \depscoht[n][\lbrack 0,p \rbrack] & \defeq &
    \left\{
    \begin{array}{l}
      \Pi q \leq n-p.\,\Pi r \leq q.\,\Pi \epsilon.\,\Pi \omega.\,\Pi \theta.\,\Pi d.                                                     \\
      \cohf[n][p][r][0][\omega][\theta](\restrf[n+2][p][q+2][\epsilon](\overline{\depscohs}(\depscoht[n][\lbrack 0,p \rbrack]))(d)) \circ \\
      \ap\restrf[n][p][r][\omega](\cohf[n+1][p][q+1][0][\epsilon][\theta]
      (\depscoht[n][p])(d))
      \circ                                                                                                                               \\
      \cohf[n][p][q][r][\epsilon][\omega](\restrf[n+2][p][0][\theta](\overline{\depscohs}(\depscoht[n][\lbrack 0,p \rbrack]))(d)) =       \\
      \ap\restrf[n][p][0][\theta](\cohf[n+1][p][q+1][r+1][\epsilon][\omega]
      (\depscoht[n][p])(d))
      \circ                                                                                                                               \\
      \cohf[n][p][q][0][\epsilon][\theta](\restrf[n+2][p][r+1][\omega](\overline{\depscohs}(\depscoht[n][\lbrack 0,p \rbrack]))(d))
      \circ                                                                                                                               \\
      \ap\restrf[n][p][q][\epsilon](\cohf[n+1][p][r][0][\omega][\theta]
      (\depscoht[n][p])(d))
    \end{array}
    \right.
  \end{array}
\end{equation*}
where $\coht$, commutativity of abstraction and application with transport, transitivity of equality and congruence, and functional extensionality relative to $\theta$ were used to go from the type\footnote{omitting the $\framep$ argument} of $(\ap \overrightarrow{\cohf[n][p][r][0][\omega][\theta]\ldots}(\ap \overrightarrow{\ap\restrf[n][p][r][\omega](\cohf[n+1][p][q+1][0][\epsilon][\theta]\ldots)}(\cohp[n][p][q][r][\epsilon][\omega](l_\theta))))$ below
\begin{equation*}
  \smfontsize
  \begin{array}{lcl}
    \begin{array}{l}
      \overrightarrow{\cohf[n][p][r][0][\omega][\theta]}\overrightarrow{\ap\restrf[n][p][r][\omega](\cohf[n+1][p][q+1][0][\epsilon][\theta])} \\
      \overrightarrow{\cohf[n][p][q][r][\epsilon][\omega]}
      (\restrp[n][p][q][\epsilon](\restrp[n+1][p][r][\omega](l_\theta)))
    \end{array} & = &
    \begin{array}{l}
      \overrightarrow{\cohf[n][p][r][0][\omega][\theta]}\overrightarrow{\ap\restrf[n][p][r][\omega](\cohf[n+1][p][q+1][0][\epsilon][\theta])} \\
      (\restrp[n][p][r][\omega](\restrp[n+1][p][q+1][\epsilon](l_\theta)))
    \end{array}
  \end{array}
\end{equation*}
to the unfolding of the type of $\cohl[n][p][q][r][\epsilon][\omega]$ below
\begin{equation*}
  \smfontsize
  \overrightarrow{\cohf[n+1][p][q+1][r+1][\epsilon][\omega]}\,\lambda \theta.\!\left(
  \!\!\!\begin{array}{l}
      \overrightarrow{\cohf[n][p][q][0][\epsilon][\theta]}(\restrp[n][p][q][\epsilon] \\
      (\overrightarrow{\cohf[n+1][p][r][0][\omega][\theta]}(\restrp[n+1][p][r][\omega](l_\theta))))
    \end{array}\!\!\!\right) \!\!=\!
  \lambda \theta.\!\left(
  \!\!\!\begin{array}{l}
      \overrightarrow{\cohf[n][p][r][0][\omega][\theta]}(\restrp[n][p][r][\omega] \\
      (\overrightarrow{\cohf[n+1][p][q+1][0][\epsilon][\theta]}(\restrp[n+1][p][q+1][\epsilon](l_\theta))))
    \end{array}\!\!\!\right)
\end{equation*}

\subsection{Defining painting coherence laws\href{https://artagnon.github.io/bonak/docs/Bonak.\%CE\%BDSet.\%CE\%BDSet.html\#mkCohPaintings}{\linkicon}\label{sec:mkcohp}}

It is convenient to introduce a notion of full dependencies at the level of the coherence conditions of dimension $2$, that combines a full $\depscoht[n][\lbrack 0,n \rbrack]$ with an extra $E''$ at level $n+3$.

\begin{equation*}
  \smfontsize
  \fulldepscohttype[n] \defeq
  \left(\begin{array}{lll}
      \depscoht[n][\lbrack 0,n \rbrack] & : & \depscohttype[n][\lbrack 0,n \rbrack]                                                                         \\
      E''                               & : & \framep[n+3][n+3](\overline{\deps}(\overline{\depscohs}(\depscoht[n][\lbrack 0,n \rbrack])))\rightarrow \HSet \\
    \end{array}\right)
\end{equation*}

We can then extend $\overline{\depscohs}$ to build $\fulldepscohs[n+1]$ from $\fulldepscoht[n]$.

\begin{equation*}
  \smfontsize
  \begin{array}{llcl}
    \overline{\fulldepscohs} & \fulldepscoht[n] & :      & \fulldepscohstype[n+1] \\
    \overline{\fulldepscohs} & \fulldepscoht[n] & \defeq & \left(
    \begin{array}{l}
        \overline{\depscohs}(\depscoht[n][\lbrack 0,n \rbrack]) \\
        E''                                                     \\
      \end{array}\right)                       \\
  \end{array}
\end{equation*}

We are then ready to define the final component of the construction, namely $\cohp[n+1][p]$ for $p \leq n+1$, by recursion on $n+1-p$.

\begin{equation*}
  \smfontsize
  \begin{array}{llcl}
    \cohp[n+1][p]                                &
    \fulldepscoht[n]                             & :      &
    \cohptype[n+1][p](\overline{\fulldepscohs}(\fulldepscoht[n]))                                \\
    \cohp[n+1][n+1][q][r][\epsilon][\omega]      &
    \fulldepscoht[n]
                                                 & \defeq & \lambda d. \lambda (l,c). l_\epsilon \\
    \cohp[n+1][p \leq n][q][r][\epsilon][\omega] &
    \fulldepscoht[n]
                                                 & \defeq & \lambda d. \lambda (l,c).
    \left(\begin{array}{l}
              \cohl[n][p][q][r][\epsilon][\omega](\depscoht[n][\lbrack 0,p-1 \rbrack])(d)(l), \\
              \cohp[n+1][p+1][q][r][\epsilon][\omega](\fulldepscoht[n])(d,l))(c)              \\
            \end{array}\right)        \\
  \end{array}
\end{equation*}

From the latter, we can deduce $\cohp[n+1][\lbrack 0,p-1 \rbrack]$ for $p \leq n+2$, by induction on $p$.

\begin{equation*}
  \smfontsize
  \begin{array}{llcl}
    \cohp[n+1][\lbrack 0,p-1 \rbrack] &
    \fulldepscoht[n]                  & :      & \cohptype[n+1][\lbrack 0,p-1 \rbrack](\fulldepscohs[n]) \\
    \cohp[n+1]                        &
    \fulldepscohs[n]                  & \defeq & \unittype                                               \\
    \cohp[n+1][\lbrack 0,p \rbrack]   &
    \fulldepscoht[n]
                                      & \defeq &
    (\cohp[n+1][\lbrack 0,p-1 \rbrack](\fulldepscoht[n]),\cohp[n+1][p](\fulldepscoht[n]))                \\
  \end{array}
\end{equation*}

\subsection{Completing the construction by induction\href{https://artagnon.github.io/bonak/docs/Bonak.\%CE\%BDSet.\%CE\%BDSet.html\#6d1cc800856804da9f9026c7c451927b}{\linkicon}}

Armed with all the constructions, we are now able to specify the whole structure on which we can reason by induction, for $D:\Xto{m}{n+1}$ given:

\begin{equation*}
  \smfontsize
  \begin{array}{lcl}
    \framep[n][\lbrack 0,n \rbrack]      & :      & \frametype[n][\lbrack 0,n \rbrack]                                     \\
    \painting[n][\lbrack 0,n \rbrack]    & :      & \paintingtype[n][\lbrack 0,n \rbrack](\framep[n][\lbrack 0,n \rbrack]) \\
    \restrf[n][\lbrack 0,n \rbrack]      & :      & \restrftype[n][\lbrack 0,n \rbrack]
    \left(
    \begin{array}{l}
        \framep[n][\lbrack 0,n \rbrack]   \\
        \painting[n][\lbrack 0,n \rbrack] \\
      \end{array}\right)                                                                                      \\
    \deps[n][\lbrack 0,n \rbrack]        & \defeq &
    \left(
    \begin{array}{l}
        \framep[n][\lbrack 0,n \rbrack]   \\
        \painting[n][\lbrack 0,n \rbrack] \\
        \restrf[n][\lbrack 0,n \rbrack]   \\
      \end{array}\right)                                                                                      \\
    \fulldeps[n](E)                      & \defeq &
    \left(
    \begin{array}{l}
        \deps[n][\lbrack 0,n \rbrack] \\
        E                             \\
      \end{array}\right)                                                                                          \\
    \restrp[n][\lbrack 0,n \rbrack](E)   & :      & \restrptype[n][\lbrack 0,n \rbrack]
    (\fulldeps[n](E))                                                                                                      \\
    \cohf[n][\lbrack 0,n \rbrack](E)     & :      & \cohftype[n][\lbrack 0,n \rbrack]
    \left(
    \begin{array}{l}
        \fulldeps[n](E)                    \\
        \restrp[n][\lbrack 0,n \rbrack](E) \\
      \end{array}\right)                                                                                     \\
    \depscohs[n][\lbrack 0,n \rbrack](E) & \defeq &
    \left(
    \begin{array}{l}
        \fulldeps[n](E)                    \\
        \restrp[n][\lbrack 0,n \rbrack](E) \\
        \cohf[n][\lbrack 0,n \rbrack](E)   \\
      \end{array}\right)                                                                                     \\
    \fulldepscohs[n](E)(E')              & \defeq &
    \left(
    \begin{array}{l}
        \depscohs[n][\lbrack 0,n \rbrack](E) \\
        E'                                   \\
      \end{array}\right)                                                                                   \\
    \cohp[n][\lbrack 0,n \rbrack](E)(E') & :      & \cohptype[n][\lbrack 0,n \rbrack]
    (\fulldepscohs[n](E)(E'))                                                                                              \\
  \end{array}
\end{equation*}

where
\begin{align*}
  E: \framep[n+1][n+1](\deps[n][\lbrack 0,n \rbrack])\rightarrow \HSet \\
  E': \framep[n+2][n+2](\overline{\deps}(\depscohs[n][\lbrack 0,n \rbrack](E)))\rightarrow \HSet
\end{align*}

To lift from level $n$ from level $n+1$, it is then enough to assume $E$ at level $n+1$. We get $\deps[n][\lbrack 0,n \rbrack]$ from the assumptions at level $n$ and we can build the following:
\begin{itemize}
  \item $\framep[n+1][\lbrack 0,n+1 \rbrack]$ comes from $\deps[n][\lbrack 0,n \rbrack]$ (section~\ref{sec:mkframe}).
  \item $\painting[n+1][\lbrack 0,n+1 \rbrack]$ comes from $\fulldeps[n](E)$ (section~\ref{sec:mkpainting}).
  \item $\restrf[n+1][\lbrack 0,n+1 \rbrack]$ comes from $\depscohs[n][\lbrack 0,n \rbrack](E)$ (section~\ref{sec:mkrestrf}).
  \item $\restrp[n+1][\lbrack 0,n+1 \rbrack](E')$ comes from $\fulldepscohs[n](E)(E')$ (section~\ref{sec:mkrestrp}).
  \item we define $\depscoht[n][\lbrack 0,n \rbrack](E)(E')$ from $\fulldepscohs[n](E)(E')$ and $\cohp[n][\lbrack 0,n \rbrack](E)(E')$.
  \item $\cohf[n+1][\lbrack 0,n+1 \rbrack](E')$ comes from $\depscoht[n][\lbrack 0,n \rbrack](E)(E')$ (section~\ref{sec:mkcohf}).
  \item we define $\fulldepscoht[n](E)(E')(E'')$, for $E'': \framep[n+3][n+3](\overline{\deps}(\overline{\depscohs}(\depscoht[n][\lbrack 0,n \rbrack]))) \rightarrow \HSet$, from $\depscoht[n][\lbrack 0,n \rbrack](E)(E')$ and $E''$.
  \item $\cohp[n+1][\lbrack 0,n+1 \rbrack](E')(E'')$ comes from $\fulldepscoht[n](E)(E')(E'')$ (section~\ref{sec:mkcohp}).
\end{itemize}

Therefore, assuming a construction of $\framep[n][n]$ from $D:\Xto{m}{n+1}$, we are able to build a $\framep[n+1][n+1]$ from $D$ extended with some $E:\framep[n+1][n+1](\deps[n][\lbrack 0,n \rbrack]) \rightarrow \HSet$. This is enough to build $\Xcomp{m}{n+1}(D)$, for $D:\Xto{m}{n+1}$, and thus $\Xp{m}$, by starting with an empty $D$, since all fields of the structure are trivially inhabited by $\kstar$ when $n = 0$.

\section{Differences between presentation and Rocq formalisation}
\begin{itemize}
  \item The formalisation is not parameterised by $n$ but instead by $k \defeq n - p$. This makes the presentation less intuitive since the dimension has to be mentally recomputed by adding $p$ to $k$, but it eases formalisation. The interesting information when building some component up to $p$ is not the dimension, but the distance from to the expected dimension, computationally speaking.
  \item The structures $\fulldeps$, $\fulldepscohs$ and $\fulldepscoht$ from the informal presentation are implemented by using a ``zipper'' structure~\cite{Huet97}: anytime we build some construction at $p$, we split $\framep[n][\lbrack 0,n \rbrack]$ into two lists constructed in opposite directions: $\framep[n][\lbrack 0,p-1 \rbrack]$ and $\framep[n][\lbrack p,n \rbrack]$, which aids in proofs of induction on $p$.
  \item In section \ref{sec:detailed-construction}, we left implicit the management of inequality proofs occurring in restrictions and coherence conditions. In practice, we stated inequalities as strict propositions~\cite{gilbert19}, or propositions whose proofs are all definitionally equal, which simplified formalisation.
  \item Rocq does not support mutual fixpoints for which one component depends on another component in its type\footnote{also termed as a recursive-recursive definition~\cite{Forsberg13}}, nor mutual fixpoints including abbreviations, as we expect in sections~\ref{sec:mkframe}, \ref{sec:mkrestrf}, and \ref{sec:mkcohf}. We address the absence of abbreviations by defining abbreviations in advance and address the absence of recursive-recursive definitions by instead recursively building the $\Sigma$-type of the two components of the recursive-recursive definition, and recovering the individual components using projections.
\end{itemize}

\section{Comparison with previous work\label{sec:previous-work}}

To address the dependency of $\restrf[n][p]$ in $\framep[n+1][p]$, and that of $\cohf[n][p]$ in $\framep[n+2][p]$ and $\restrf[n+1][p]$ without having to mix the specification of the induction hypothesis and the construction of the next step, the previous work~\cite{hr25} was assuming all of $\framep[n][p]$, $\framep[n+1][p]$, $\framep[n+2][p]$, $\restrf[n][p]$, $\restrf[n+1][p]$, and $\cohf[n][p]$, together with propositional equations defining $\framep[n+1][p]$, $\framep[n+2][p]$, and $\restrf[n+1][p]$. The induction step then constructed $\framep[n+3][p]$, $\restrf[n+2][p]$ and $\cohf[n+1][p]$ and inherited $\framep[n+1][p]$, $\framep[n+2][p]$ and $\restrf[n+1][p]$ from the induction hypothesis. The current work can be seen as the result of analysing how dependencies are finely organised so that those propositional equations are replaced by definitions of $\framep[n+1][p]$, $\framep[n+2][p]$ and $\restrf[n+1][p]$. In other words, in the previous work, the induction hypothesis combines assumptions of the form $\framep[n+1][p] : \HSet$ with equations of the form $\framep[n+1][0] = \unittype$ and $\framep[n+1][p+1] = \Sigma d:\framep[n+1][p].\Pi \epsilon.\painting[n][p](\restrf[n][p][0][\epsilon](d))$. These are replaced by a \emph{definition} of $\framep[n+1][p]$. And similarly for $\framep[n+2][p]$, $\restrf[n+1][p]$, as well as for $\painting[n+1][p]$, $\painting[n+2][p]$, $\restrp[n+1][p]$, in the current work.

\section{Perspectives}

Our construction shows a pattern: start from $\frametype$, define $\paintingtype$, mutually define $\framep$ and $\restrftype$, define $\painting$, define $\restrptype$, mutually define $\restrf$ and $\cohftype$, define $\restrp$, define $\cohptype$, mutually define $\cohf$ and $\cohttype$, define $\cohptype$. On $\HSet$, $\cohttype$ is trivially provable and the construction stops here. If we were working in an untruncated universe $U$, the construction would continue recursively: prove the coherence at dimension $i$ on frames mutually with the type of coherence on frames at dimension $i+1$, prove the coherence on paintings at dimension $i$, define the type of coherence on paintings at dimension $i+1$. Such an infinite process can again be represented using streams, so that from a stream at level $n$, a stream at level $n+1$ can be built, as in section~\ref{sec:detailed-construction}. This would solve the semi-simplicial type problem~\cite{UFwiki2012}.

\bibliographystyle{plainurl}
\bibliography{paper}

\end{document}